\documentclass[prl,twocolumn,superscriptaddress,floatfix,noeprint,longbibliography]{revtex4-2}

\usepackage{graphicx,tabularx,makecell}
\usepackage{amsmath,amsfonts,amssymb}
\usepackage{mathtools}
\usepackage{empheq}
\usepackage{bm,dsfont}
\usepackage[dvipsnames]{xcolor}
\usepackage[colorlinks=true, citecolor=Green, linkcolor=BrickRed, urlcolor=NavyBlue]{hyperref}
\usepackage[all]{hypcap}
\usepackage[T5,T1]{fontenc}

\graphicspath{{figures/}}

\DeclareMathOperator{\sgn}{sgn}

\begin{document}

\title{Berry Curvature Spectroscopy from Bloch Oscillations}

\author{Christophe De Beule}
\affiliation{Department of Physics and Astronomy, University of Pennsylvania, Philadelphia PA 19104}
\affiliation{Department of Physics and Materials Science, University of Luxembourg, L-1511 Luxembourg, Luxembourg}
\author{E. J. Mele}
\affiliation{Department of Physics and Astronomy, University of Pennsylvania, Philadelphia PA 19104}

\date{\today}

\begin{abstract}
Artificial crystals such as moir\'e superlattices can have a real-space periodicity  much larger than the underlying atomic scale. This facilitates the presence of Bloch oscillations in the presence of a static electric field. We demonstrate that the optical response of such a system, when dressed with a static field, becomes resonant at the frequencies of Bloch oscillations, which are in the THz regime when the lattice constant is of the order of $10 \; \text{nm}$. In particular, we show within a semiclassical band-projected theory that resonances in the dressed Hall conductivity are proportional to the lattice Fourier components of the Berry curvature. We illustrate our results with a low-energy model on an effective honeycomb lattice.
\end{abstract}

\maketitle

Nonlinear optical responses are becoming an increasingly important tool to investigate the spectral and geometric properties of electron Bloch bands in low-dimensional materials \cite{Morimoto2016,wu2017giant,Ahn2022}. In particular, the nonlinear Hall effect \cite{Sodemann2015} which probes multipoles of the Berry curvature of the band at successive orders in the driving field \cite{Zhang2023}. Since time-reversal symmetry only precludes odd powers of the field in the Hall response, nonlinear responses allow one to study the momentum-space distribution of the Berry curvature even in systems with time-reversal symmetry. Recently, the advent of moir\'e \cite{Andrei2020,Andrei2021,Mak2022} and other two-dimensional (2D) artificial crystals \cite{Tsu2005,Forsythe2018,Mao2020} has opened up the prospect of studying responses at nonperturbative order in the driving field \cite{Fahimniya2021,Phong2022b,debeule2023rose}. These systems can host spectrally isolated and flattened minibands, and nonlinear responses have already been used to study their properties \cite{Pantaleon2021, He2021, Sinha2022, Chakraborty2022, Zhang2022, Pantaleon2022, Duan2022, Zhong2023}. Moreover, because the real-space periodicity of these systems can be much larger than the underlying atomic scale periodicity, with lattice constants ranging between $1$--$100$~nm, the momentum space Brillouin zone (BZ) is relatively small. Under an applied electric field, it therefore becomes possible for an electron to traverse the \emph{entire} zone, i.e., perform a full Bloch oscillation \cite{Bloch1929,LEO1992943}, before relaxing back to equilibrium by scattering. To quantify this regime, consider an applied uniform electric field of the form
\begin{equation} \label{eq:Efield}
    \bm E(t) = \bm E_0 + \bm E_1(t),
\end{equation}
which has a static component $\bm E_0 = E_0 \left( \cos \theta_0,\, \sin \theta_0 \right)$ and an oscillating component $\bm E_1(t)$. The latter acts as a weak probe for the system that is \emph{dressed} by the static field. Here, the nonperturbative regime is defined by the condition $\omega_B \tau \gg 1$ \cite{Fahimniya2021,Phong2022b,debeule2023rose} where $\omega_B = eE_0 L / \hbar$ is the Bloch frequency, i.e.\ the characteristic frequency of Bloch oscillations, and $\tau$ is the momentum-relaxation time with $L$ the lattice constant. If we estimate $\tau = 1 \; \text{ps}$ we find that 
$\omega_B \tau \approx \tfrac{1.5E_0}{\text{kV/cm}} \tfrac{L}{10 \; \text{nm}}$ such that $\omega_B \tau$ can become large in artificial crystals for reasonable field strengths \cite{Fahimniya2021,Phong2022b}. 
\begin{figure}
    \centering
    \includegraphics[width=\linewidth]{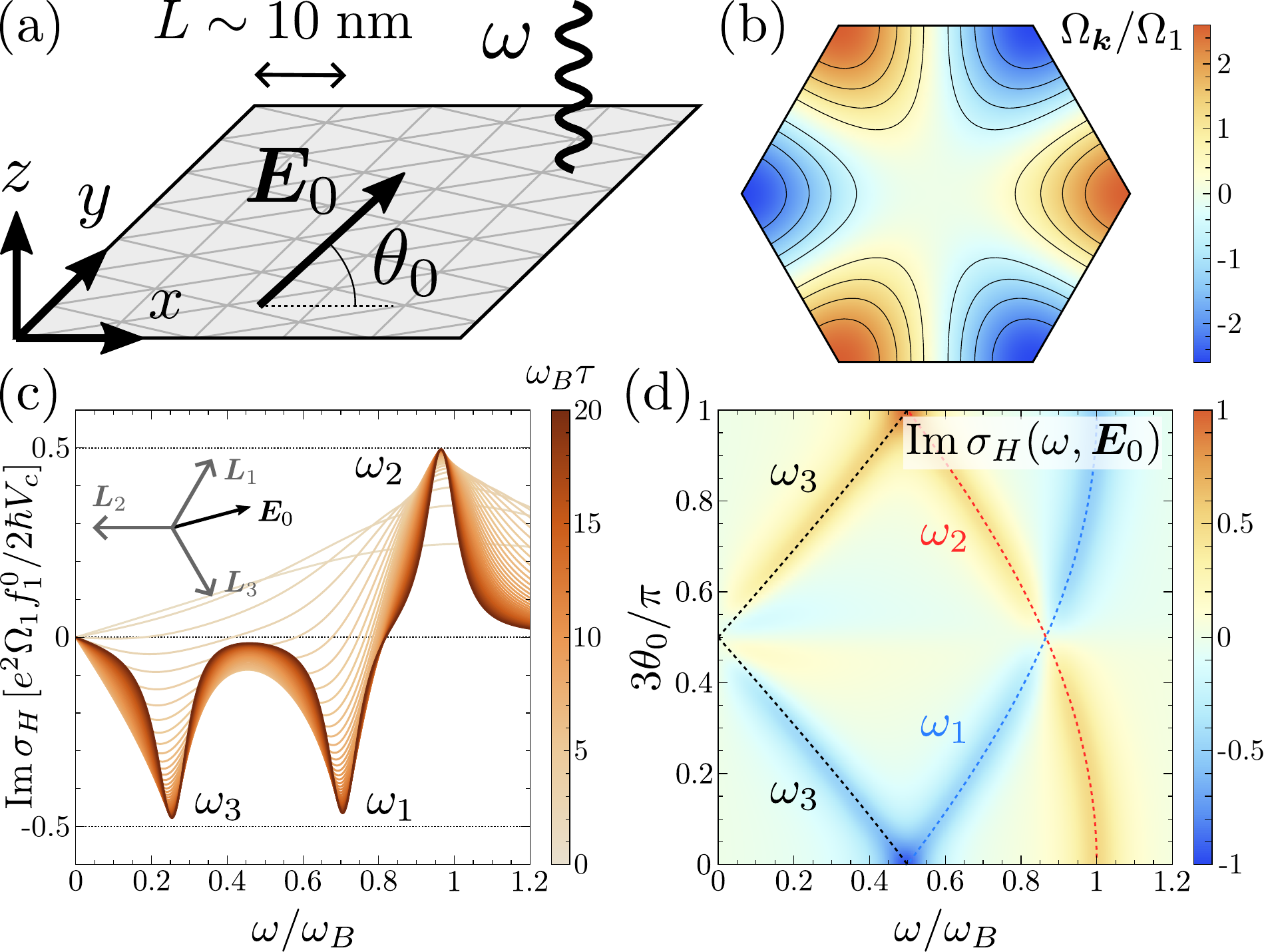}
    \caption{(a) A 2D artificial crystal (e.g., a moir\'e) subjected to a static uniform in-plane electric field $\bm E_0$ and probed by monochromatic light of frequency $\omega$. (b) Berry curvature $\Omega_{\bm k}$ in the first-shell approximation for a system with $D_3$ or $C_{3v}$ symmetry. (c) Imaginary part of the dressed optical Hall conductivity $\sigma_H(\omega,\bm E_0)$ for the Berry curvature shown in (b) as a function of $\omega/\omega_B$ for $\theta_0 = 15^\circ$ and different values of $\omega_B \tau$. (d) $\text{Im} \, \sigma_H$ in units $e^2 \Omega_1 f_1^0 / 2 \hbar V_c$ for $\omega_B \tau = 15$ as a function of the frequency and the field direction $\theta_0$. The resonant frequencies for the first shell $\omega_n^{(1)} = |e \bm E_0 \cdot \bm L_n/\hbar|$ are indicated.}
    \label{fig:fig1}
\end{figure}

In this work, we study the dressed time-dependent response of time-reversal-invariant 2D artificial lattices with lattice constants $L \sim 10 \; \text{nm}$, that are subjected to a uniform electric field of the form given in Eq.\ \eqref{eq:Efield}. This setup is illustrated in Fig.\ \ref{fig:fig1}(a). When the static field is in the regime of Bloch oscillations, we find an optical response, linear in the \emph{oscillating} component, that is resonant at the Bloch frequencies. For the studied systems, the latter are on the order of $5$--$10 \; \text{THz}$. Moreover, we show that the peak heights of the resonances in the dressed optical \emph{Hall} conductivity are proportional to the Fourier components of the Berry curvature. Hence our approach is in some sense dual to probing the momentum-space distribution of the Berry curvature via its multipoles at successive harmonics  \cite{Luu2018} and complementary to other methods that study orbital moments with circular dichroism \cite{Schuler2020}. In contrast, in our proposal, all information on the Berry curvature is contained in the dressed \emph{linear} optical response and contributions from different Fourier components can be favored by varying the direction of the static field.

\textcolor{NavyBlue}{\emph{Semiclassical theory.}} Our starting point is the band-projected semiclassical theory of electron dynamics for a 2D crystal in a uniform electric field $\bm E(t)$. The equations of motion for the central position and crystal momentum of a wave packet constructed from the Bloch states of an energy band $\varepsilon_{n\bm k}$ are given by \cite{Chang1995,Sundaram1999}
\begin{align}
    \hbar \dot{\bm r}_{n\bm k} & = \nabla_{\bm k} \varepsilon_{n\bm k} - \hbar \dot{\bm k} \times \Omega_{n\bm k} \hat z, \label{eq:eom1} \\
    \hbar \dot{\bm k} & = -e \bm E(t), \label{eq:eom2}
\end{align}
where $-e$ is the electron charge and $\Omega_{n\bm k} = - 2 \, \text{Im} \, \langle \partial_{k_x} u_{n\bm k} | \partial_{k_y} u_{n\bm k} \rangle_\text{cell}$ is the Berry curvature \footnote{To construct a wave packet $\left| W(t) \right> = \int_{\bm k} c_{\bm k}(t) \left| \Psi_{\bm k} \right>$, the Bloch states $\left| \Psi_{\bm k} \right> = e^{i \bm k \cdot \hat{\bm r}} \left| u_{\bm k} \right>$ should be smooth on the BZ torus, i.e., $\left| \Psi_{\bm k + \bm G} \right> = \left| \Psi_{\bm k} \right>$ with $\bm G$ a reciprocal lattice vector. This is periodic gauge \cite{Vanderbilt2018} and yields $\left| u_{\bm k + \bm G} \right> = e^{-i\bm G \cdot \hat{\bm r}} \left| u_{\bm k} \right>$, in contrast to $\left| \tilde u_{\bm k + \bm G} \right> = \left| \tilde u_{\bm k} \right>$ for which the Bloch Hamiltonian is periodic (Bloch form). The Berry curvature is generally different in both gauges.}. The band-projected theory holds as long as interband transitions can be neglected. These can arise both from optical transitions and electric breakdown (Zener tunneling) \cite{AshcroftMermin}. The former are absent for frequencies below the energy gap to the other energy bands $\varepsilon_\text{gap}$, while the absence of the latter can be estimated by the condition that $\varepsilon_\text{gap}^2/\varepsilon_\text{width} \gg eE_0L$ where $\varepsilon_\text{width}$ is the bandwidth. Hence, we consider the intermediate regime $\hbar/\tau \ll eE_0L \ll \varepsilon_\text{gap}^2/\varepsilon_\text{width}$ \cite{Fahimniya2021,Phong2022b,debeule2023rose}.

In the following, we drop the band index $n$ since we consider a single band. The current is then given by 
\begin{equation} \label{eq:current}
    \bm j(t) = -e \int_{\bm k} \, \dot{\bm r}_{\bm k}(t) f_{\bm k}(t),
\end{equation}
with $\int_{\bm k} = \int_\text{BZ} d^2\bm k / (2 \pi )^2$ and where $f_{\bm k}(t)$ is the nonequilibrium occupation of the electrons in the band. The latter is obtained from the Boltzmann transport equation in the relaxation-time approximation:
\begin{equation} \label{eq:boltzmann}
    \tau \partial_t f_{\bm k} - \frac{e\tau}{\hbar} \bm E(t) \cdot \nabla_{\bm k} f_{\bm k} = f_{\bm k}^0 - f_{\bm k},
\end{equation}
where $\tau$ is the momentum-relaxation time and $f_{\bm k}^0 = n_F(\varepsilon_{\bm k} - \mu)$ with $n_F$ the Fermi function and $\mu$ the chemical potential. Because the system has translational symmetry, the occupation function is periodic in momentum space: $f_{\bm k} = \sum_{\bm R} f_{\bm R} e^{i \bm k \cdot \bm R}$ where the sum runs over lattice vectors $\bm R$ with $f_{\bm R} = V_c \int_{\bm k} f_{\bm k} e^{-i\bm k \cdot \bm R}$. Plugging this expansion in Eq.\ \eqref{eq:boltzmann} we obtain an ordinary differential equation with the steady-state solution \cite{Mikhailov2017}
\begin{equation} \label{eq:fR}
    f_{\bm R}(t) = f^0_{\bm R} \int_0^\infty ds \, e^{-s} \exp \left[ \frac{ie}{\hbar} \int_{t-s\tau}^t dt' \, \bm E(t') \cdot \bm R \right],
\end{equation}
as shown in the Supplemental Material (SM) \cite{sm}. The occupation $f_{\bm k}$ is thus given by a weighted sum of displaced Fermi functions where the drift due to the electric field is determined by the accumulated momentum between collisions at time $t-s\tau$ and time $t$. Here the exponential weight $e^{-s}$ reflects the fact that scattering is modeled as a Poisson process.

The current in Eq.\ \eqref{eq:current} can be decomposed into two terms as $\bm j(t) = \bm j_\text{Bloch}(t) + \bm j_\text{geom}(t)$ where
\begin{align}
    \bm j_\text{Bloch}(t) & = \frac{ie}{\hbar V_c} \sum_{\bm R} \bm R \varepsilon_{-\bm R} f_{\bm R}(t), \\
    \bm j_\text{geom}(t) & = \hat z \times \frac{e^2}{\hbar V_c}  \sum_{\bm R} \Omega_{-\bm R} \bm E(t) f_{\bm R}(t),
\end{align}
where $V_c$ is the unit cell area and we made use of the expansions of the band dispersion and the Berry curvature, as well as $V_c \int_{\bm k} e^{i \bm k \cdot \bm R} = \delta_{\bm R,\bm 0}$. The Bloch current $\bm j_\text{Bloch}$ originates from the band dispersion while the geometric current $\bm j_\text{geom}$ originates from the anomalous velocity due to the Berry curvature in Eq.\ \eqref{eq:eom1}.

\textcolor{NavyBlue}{\emph{Dressed optical conductivity.}}  We now consider probing the system by monochromatic light of frequency $\omega$ at normal incidence. In the electric-dipole approximation, the electric field of the light can be written as
\begin{equation}
    \bm E_1(t) = \bm{\mathcal E}_1 e^{i\omega t} + \bm{\mathcal E}_1^* e^{-i\omega t},
\end{equation}
where $\bm{\mathcal E}_1 \in \mathds C^2$ gives the amplitude and polarization. To investigate the response at frequency $\omega$, we expand each lattice Fourier component of the distribution function in its frequency components. We have $f_{\bm R}(t) = \sum_{m=-\infty}^\infty f_{\bm R,m} e^{im\omega t}$ where $f_{\bm R,m} = ( \omega / 2\pi ) \int_0^{2\pi/\omega} dt \, f_{\bm R}(t) e^{-im\omega t}$ with $f_{\bm R,-m} = f_{-\bm R,m}^*$. The frequency components of the currents become
\begin{align}
    \bm j_\text{Bloch}^{(m)} & = \frac{ie}{\hbar V_c} \sum_{\bm R} \bm R \, \varepsilon_{-\bm R} f_{\bm R,m}, \\
    \begin{split}
    \bm j_\text{geom}^{(m)} & = \hat z \times \frac{e^2}{\hbar V_c} \sum_{\bm R} \Omega_{-\bm R} \\
    & \cdot \left( \bm E_0 f_{\bm R,m} + \bm{\mathcal E}_1 \, f_{\bm R,m-1} + \bm{\mathcal E}_1^* \, f_{\bm R,m+1} \right).
    \end{split}
\end{align}
Since we are interested in the linear response \emph{dressed} by the static part of the field, we expand Eq.\ \eqref{eq:fR} in orders of $|e \bm{\mathcal E}_1 \cdot \bm R / \hbar \omega|$ while retaining all orders in $\bm E_0$. Up to first order, the only nonzero terms are given by
\begin{align}
    f_{\bm R,0} & = \frac{f^0_{\bm R}}{1 - i \omega_{\bm R} \tau}, \\
    f_{\bm R,1} & = \frac{f^0_{\bm R}}{1 - i \omega_{\bm R} \tau} \frac{e \bm{\mathcal E}_1 \cdot \bm R / \hbar}{\omega - \omega_{\bm R} - \frac{i}{\tau}} = f_{-\bm R,-1}^*,
\end{align}
with $\omega_{\bm R} = e \bm E_0 \cdot \bm R / \hbar$.
\begin{figure}
    \centering
    \includegraphics[width=\linewidth]{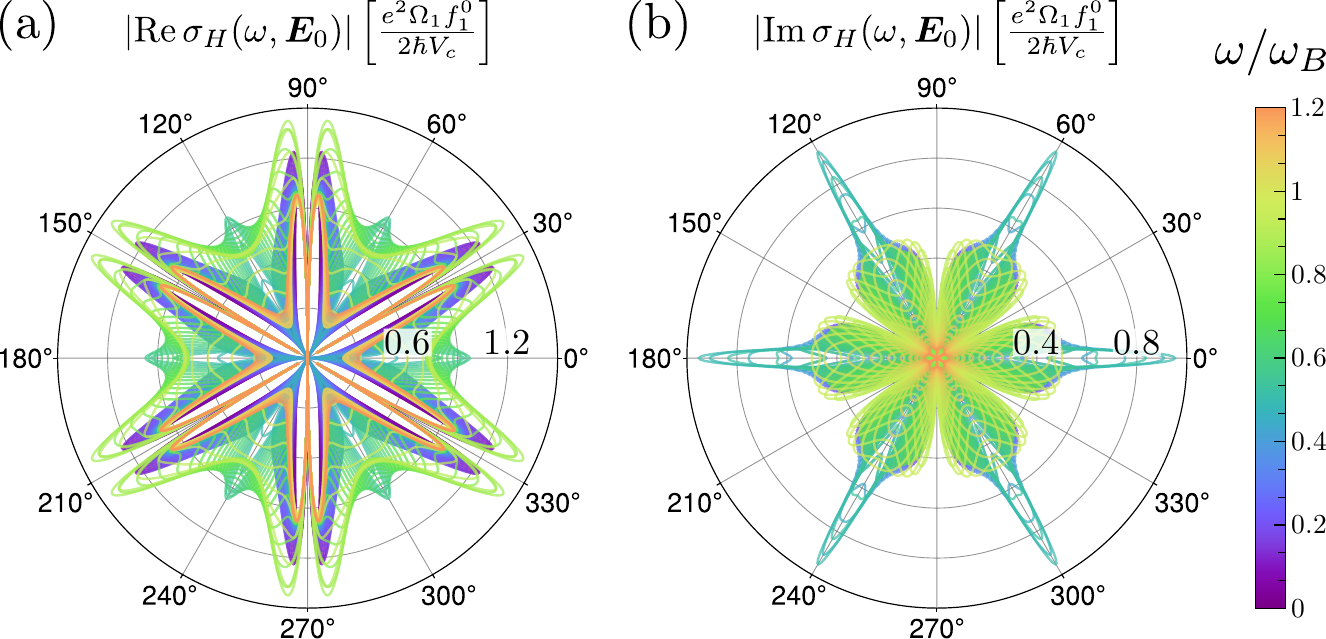}
    \caption{Roses for the real (a) and imaginary (b) part of the dressed optical Hall conductivity $\sigma_H(\omega,\bm E_0)$ for the Berry curvature shown in Fig.\ \ref{fig:fig1}(b) with $\omega_B \tau = 15$. The angle corresponds to the direction of the static electric field $\theta_0$ and the color scale gives the frequency $\omega$ of the oscillating field.}
    \label{fig:fig2}
\end{figure}
The response at frequency $\omega$ can then be written as $j_a^{(1)} = \sigma_{ab} \mathcal E_{1b}$ where $a,b = x,y$ and summation over repeated indices is implied. This leads us to the main result of this work: the \emph{dressed} optical conductivity
\begin{equation} \label{eq:sigma}
    \begin{aligned}
     & \sigma_{ab}(\omega,\bm E_0) = \frac{ie^2}{\hbar^2 V_c} \sum_{\bm R} \frac{R_a R_b \varepsilon_{-\bm R} f_{\bm R}^0}{\left( 1 - i \omega_{\bm R} \tau \right) \left( \omega - \omega_{\bm R} - \frac{i}{\tau} \right)} \\
    & - \frac{e^2}{\hbar V_c} \sum_{\bm R} \frac{\Omega_{-\bm R} f_{\bm R}^0}{1 - i \omega_{\bm R} \tau} \left[ \epsilon_{ab}  + \frac{e \epsilon_{ac} E_{0c} R_b / \hbar}{\omega - \omega_{\bm R} - \frac{i}{\tau}} \right],
    \end{aligned}
\end{equation}
where $\epsilon_{ab}$ is the permutation symbol and $\sigma_{ab}(\omega,\bm E_0)^* = \sigma_{ab}(-\omega,\bm E_0)$ such that the real (imaginary) part is even (odd) in $\omega$. As a check, we undress the conductivity by setting $E_0 = 0$. In this case, the two terms in Eq.\ \eqref{eq:sigma} reduce to the Drude and anomalous Hall conductivity, respectively. Importantly, the dressed linear Hall response does \emph{not} vanish when time-reversal symmetry is conserved, because it is effectively a compound nonlinear response in the fields $\bm E_0$ and $\bm E_1(t)$. 

Let us now focus on the case where $\bm E_0$ is finite and consider the dressed longitudinal $\sigma_L = \delta_{ab} \sigma_{ab}/2$ and Hall $\sigma_H = \epsilon_{ab} \sigma_{ab}/2$ conductivities, which transform as a scalar and pseudoscalar, respectively \footnote{Note that one also has to transform $\bm E_0$ such that $\sigma_H$ is nonzero even in the presence of mirror symmetry.}. We obtain
\begin{align}
    \sigma_L & = \frac{ie^2}{2\hbar^2 V_c} \sum_{\bm R} \frac{R^2 \varepsilon_{-\bm R} f^0_{\bm R}}{\left( 1 - i \omega_{\bm R} \tau \right) \left( \omega - \omega_{\bm R} - \frac{i}{\tau} \right)}, \\
    \sigma_H & = -\frac{e^2}{\hbar V_c} \sum_{\bm R} \frac{\Omega_{-\bm R} f^0_{\bm R}}{1 - i \omega_{\bm R} \tau} \left( 1 + \frac{1}{2} \frac{\omega_{\bm R}}{\omega - \omega_{\bm R} - \frac{i}{\tau}} \right),
\end{align}
which for $\omega_B \tau \gg 1$ simplify to
\begin{align}
    \sigma_L(\omega,\bm E_0) & = -\frac{e^2}{h} \frac{\pi}{\tau V_c} \sum_{\bm R} \frac{R^2 \varepsilon_{-\bm R} f^0_{\bm R}}{\hbar \omega_{\bm R} \left( \omega - \omega_{\bm R} - \frac{i}{\tau} \right)}, \label{eq:sigmaL} \\
    \sigma_H(\omega,\bm E_0) & = -\frac{e^2}{h} \frac{\pi}{\tau V_c} \sum_{\bm R} \frac{i\Omega_{-\bm R} f^0_{\bm R}}{\omega - \omega_{\bm R} - \frac{i}{\tau}}.
\end{align}
For crystals with time-reversal symmetry, the band dispersion (Berry curvature) is an even (odd) function of momentum, such that $\varepsilon_{\bm R}$ and $f_{\bm R}^0$ are real, while $\Omega_{\bm R}$ is imaginary. In this case, and for $\omega_B \tau \gg 1$, we see that $\text{Im} \, \sigma_L$ and $\text{Im} \, \sigma_H$ are given by a series of Lorentzians centered at the Bloch frequencies $\omega_{\bm R}$. The height of these resonances is proportional to $\varepsilon_{\bm R}$ and $\Omega_{\bm R}$, respectively, and independent of the relaxation time $\tau$. Conversely, the real part of the dressed conductivity vanishes at resonance. 
Hence $\sigma_L$ is purely reactive while $\sigma_H$ is purely absorptive at Bloch resonance. For linearly polarized light, the system does not dissipate, since it is essentially collisionless on the time scale set by Bloch oscillations for $\omega_B \tau \gg 1$. However, for circularly polarized light the Hall response couples dissipatively via $\text{Im} \, \sigma_H$ since it lags in phase by a quarter cycle (see also SM).

These results can thus potentially be used to map out the distribution of the Berry curvature in systems with time-reversal symmetry by measuring the resonances in the dressed optical Hall conductivity in the nonperturbative regime where $\omega_B \tau \gg 1$.

\textcolor{NavyBlue}{\emph{First-shell approximation.}} It is instructive to first evaluate the dressed optical conductivity by only taking into account the leading-order terms in the sum over the lattice vectors. For concreteness, we consider a system with point group $D_3$ or $C_{3v}$ which lacks inversion or $\mathcal C_{2z}$ rotation symmetry. In this case, the Berry curvature is generally nonzero even though the Chern number of the band vanishes. In the first-shell approximation, we only take into account the shortest nonzero lattice vectors such that $\varepsilon_{\bm k} = \varepsilon_1 \sum_{n=1}^3 \cos( \bm k \cdot \bm L_n )$ up to a constant and $\Omega_{\bm k} = \Omega_1 \sum_{n=1}^3 \sin( \bm k \cdot \bm L_n )$ where $\varepsilon_1$ and $\Omega_1$ are real parameters that depend on the details of the system, and $\bm L_1 = L ( 1/2, \sqrt{3}/2 )$, $\bm L_2 = (-L,0)$, and $\bm L_3 = -(\bm L_1 + \bm L_2)$ are related by $\mathcal C_{3z}$ rotation symmetry \cite{Phong2022b,debeule2023rose}.

The imaginary part of the dressed optical Hall conductivity is shown in Fig.\ \ref{fig:fig1}(c) as a function of $\omega$ for $\theta_0 = 15^\circ$ and different values of $\omega_B \tau$. There are three resonances in this case because the first coordination shell supports three Bloch frequencies $\omega_n^{(1)} = |e \bm E_0 \cdot \bm L_n|$ which are nondegenerate for  general $\theta_0$. The height of these resonances is approximately equal due to $\mathcal C_{3z}$ and time-reversal symmetry and saturates to $e^2 \Omega_1 f_1^0 / 2 \hbar V_c$ in the limit $\omega_B \tau \gg 1$, where $f_1^0 = f_{\bm R=\bm L_n}^0$. Notice that the resonances are only well-defined for $\omega_B \tau \gtrsim 10$. The dependence on the direction of the static field is shown in Fig.\ \ref{fig:fig1}(d). Here we show $\text{Im} \, \sigma_H$ for $\omega_B \tau = 15$ as a function of $\omega$ and $\theta_0$. As we rotate the static field, resonances move along the curves $\omega = \pm \omega_B \cos(\theta_0 - \theta_n)$ with $\theta_n = \{\pi/3, \pi, -\pi/3\}$. For the special case $\theta_0 = m\pi/3$ ($m \in \mathbb Z$) two Bloch frequencies coincide and the peaks are doubled. On the contrary, for $\theta_0 = (2m+1)\pi/6$ the response vanishes due to $\mathcal M_x$ ($x \mapsto -x$) mirror symmetry. These features can also be seen in the rose plots of Fig.\ \ref{fig:fig2}. Here we clearly see that the strongest resonance occurs when two lattice vectors have the same projection along the static field. Away from these directions, the resonance splits into two peaks that shift to higher and lower frequencies.

\textcolor{NavyBlue}{\emph{Low-energy model.}} Going beyond the first-shell approximation, we now consider a low-energy model defined on an effective honeycomb lattice with one orbital per site, and with nearest-neighbor hopping amplitude $t>0$ and a sublattice-staggering potential $m$. The Bloch Hamiltonian is given by
\begin{gather}
    \mathcal H(\bm k) = \bm d(\bm k) \cdot \bm \sigma, \\
    \bm d(\bm k) = \left( -t \, \text{Re} \, g_{\bm k}, \, -t \, \text{Im} \, g_{\bm k}, \, m \right),
\end{gather}
where $\bm \sigma = \left( \sigma_x , \sigma_y, \sigma_z \right)$ are the Pauli matrices and $g_{\bm k} = e^{-i \bm k \cdot \bm \tau} \left[ 1 + e^{i\bm k \cdot \bm L_1} + e^{i \bm k \cdot \left( \bm L_1 + \bm L_2 \right)} \right]$ with $\bm \tau = L \hat y / \sqrt{3}$ the relative separation of the two sublattices. Note that we work in periodic gauge for which the semiclassical equation given in Eq.\ \eqref{eq:eom1} is valid \cite{Sundaram1999,Vanderbilt2018}. This model has time-reversal symmetry with point group $C_{3v}$ generated by $\mathcal C_{3z}$ and $\mathcal M_x$, and can be seen as a minimal low-energy model for moir\'es such as hBN-aligned twisted bilayer graphene \cite{Zhang2019,Lewandowski2019} or twisted double bilayer graphene \cite{Koshino2019,Chebrolu2019}, as well as other systems belonging to the same symmetry class such as periodically-buckled graphene with a $C_{3v}$ height profile \cite{Milovanovic2020,Phong2022,Gao2022,debeule2023rose}. 
\begin{figure}
    \centering
    \includegraphics[width=\linewidth]{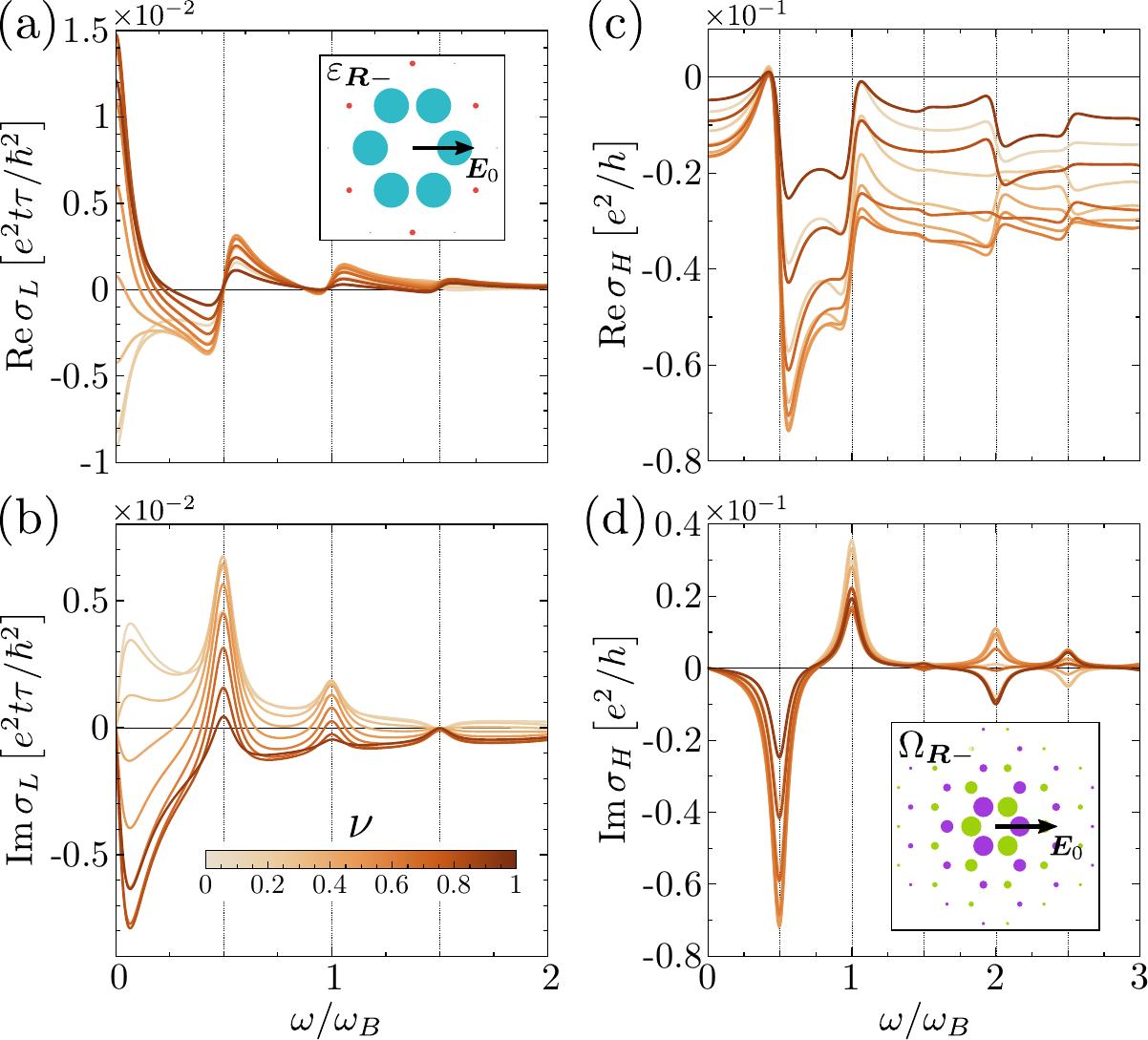}
    \caption{Dressed optical conductivities $\sigma_L(\omega, \bm E_0)$ and $\sigma_H(\omega, \bm E_0)$ for the valence band of the two-band model with $m/t = 0.5$ where $\omega_B \tau = 15$, $\theta_0 = 0^\circ$, and $k_B T / t = 0.004$. The color scale gives the filling $\nu \in [0.1,0.9]$ in $0.1$ increments [see inset of (b)]. (a,\,b) Real and imaginary part of the longitudinal conductivity. (c,\,d) Real and imaginary part of the Hall conductivity. Dashed vertical lines give the position of the resonances $\omega_{\bm R}$ and the inset in (a) and (d) shows the relative magnitude and phase of $\varepsilon_{\bm R-}$ and $\Omega_{\bm R-}$, respectively.}
    \label{fig:fig3}
\end{figure}

The model gives two energy bands $\varepsilon_{\bm k\pm} = \pm |\bm d(\bm k)|$ that are separated by a gap $|2m|$ at the zone corners. Because $\mathcal C_{2z}$ symmetry is broken by the sublattice potential, the Berry curvature is nonzero and given by 
\begin{equation}
    \Omega_{\bm k\pm} = \pm \frac{m t^2 V_c}{6 |\bm d(\bm k)|^3}  \sum_{n=1}^3 \sin \left( \bm k \cdot \bm L_n \right),
\end{equation}
with $V_c = \sqrt{3} \, L^2 / 2$. In the limit $|m/t| \gg 1$, we have $|\bm d(\bm k)| \simeq |m|$ and the first shell dominates with $\Omega_1 = \pm \sgn(m) V_c \, t^2 / 6m^2$. However, in general many shells contribute, as illustrated in Fig.\ \ref{fig:fig3} where we show $\sigma_L(\omega,\bm E_0)$ in panels (a) and (b), and $\sigma_H(\omega,\bm E_0)$ in panels (c) and (d) for $m/t=0.5$ and different fillings $\nu$ of the valence band. Here the static field lies along the $x$ direction and $k_B T / t \ll 1$. Note $\sigma_L$ decays faster with frequency than $\sigma_H$ because the first shell of the dispersion is dominant [see inset of Fig.\ \ref{fig:fig3}(a)] and because of the additional factor of $1/\omega$ in Eq.\ \eqref{eq:sigmaL}. Note that the filling $\nu$ enters only through the Fourier components of the Fermi function $f_{\bm R}^0$ which modulate the height of the peaks in the imaginary part of the conductivities and can change sign as a function of $\nu$, see Fig.\ \ref{fig:fig3}(d). 
 
In conclusion, we developed a band-projected semiclassical theory for the optical response of an artificial crystal, such as a moir\'e material, that is dressed by a uniform static field. When the static field is sufficiently strong, achieved for field strengths of order $10 \; \text{kV/cm}$ for a lattice constant of order $10 \; \text{nm}$, the dressed system becomes resonant at the Bloch frequencies which are in the $10 \; \text{THz}$ regime. We quantified this effect by defining a dressed optical conductivity whose imaginary part displays resonant peaks, while the real part vanishes at resonance. In particular, the height of the resonances in the optical Hall conductivity probe the lattice Fourier components of the Berry curvature and are independent of the relaxation time. One thus obtains an intrinsic probe of the quantum geometry of the band by resonantly coupling light to Bloch oscillations. The dressed optical conductivity can for example be obtained from THz Faraday rotation and ellipticity spectroscopy measurements \cite{Spielman1994,Shimano2011}. In contrast to probes of the Berry curvature multipoles, such at the rectified second-order response involving the Berry curvature dipole \cite{Sodemann2015}, our proposal works at linear order in the optical response, and works best for a smooth Berry curvature dominated by the first coordination shell whose lowest multipoles are zero or small. Moreover, by changing the in-plane direction of the static field, one can tune contributions from different lattice vectors. This work thus provides a novel route to probe the Berry curvature in time-reversal symmetric moir\'e and other artificial crystals which have a large real-space periodicity.

\let\oldaddcontentsline\addcontentsline 
\renewcommand{\addcontentsline}[3]{} 
\begin{acknowledgments} 
We thank V.~T.\ Phong for discussions. This research was funded in whole, or in part, by the Luxembourg National Research Fund (FNR) (project No.\ 16515716). C.~D.~B.\ and E.~J.~M.\ are supported by the Department of Energy under grant DE-FG02-84ER45118.
\end{acknowledgments}
\bibliography{references}
\let\addcontentsline\oldaddcontentsline 
\newpage\hbox{}\thispagestyle{empty}\newpage

\onecolumngrid
\begin{center}
\textbf{\large Supplemental Material for ``Berry Curvature Spectroscopy from Bloch Oscillations''}
\end{center}
\setcounter{equation}{0}
\setcounter{figure}{0}
\setcounter{table}{0}
\setcounter{page}{1}
\makeatletter
\renewcommand{\thepage}{S\arabic{page}}
\renewcommand{\theequation}{S\arabic{equation}}
\renewcommand{\thefigure}{S\arabic{figure}}

\tableofcontents

\section{Semiclassical model of electron dynamics}

The semiclassical equations of motion for an electron in a two-dimensional (2D) crystal, occupying an energy band with dispersion $\varepsilon_{n\bm k}$ subjected to a uniform electric field $\bm E(t)$ are given by \cite{Chang1995,Sundaram1999}
\begin{subequations}
\begin{align}
    \hbar \dot{\bm r}_{n\bm k} & = \nabla_{\bm k} \varepsilon_{n\bm k} - \hbar \dot{\bm k} \times \bm \Omega_{n\bm k}, \\
    \hbar \dot{\bm k} & = -e \bm E(t),
\end{align}
\end{subequations}
where the dot stands for the time derivative $d/dt$. Here $e>0$ is the elementary charge, $n$ is the band index, and $\bm \Omega_{n\bm k} = \Omega_{n\bm k} \hat z$ is the Berry curvature. The latter is defined as
\begin{equation}
    \Omega_{n\bm k} = i \left( \left\langle \frac{\partial u_{n\bm k}}{\partial k_x} \bigg\rvert \frac{\partial u_{n\bm k}}{\partial k_y} \right\rangle_\text{cell} - \left\langle \frac{\partial u_{n\bm k}}{\partial k_y} \bigg\rvert \frac{\partial u_{n\bm k}}{\partial k_x} \right\rangle_\text{cell} \right),
\end{equation}
where $u_{n\bm k}(\bm r)$ are cell-periodic Bloch functions in periodic gauge, $u_{n,\bm k+\bm G}(\bm r) = e^{-i \bm G \cdot \bm r} u_{n\bm k}(\bm r)$ with $\bm G$ a reciprocal lattice vector, and $\langle u_{n\bm k} | u_{m\bm k} \rangle_\text{cell} = \delta_{nm}$. 

In the following, we consider a single band and omit the band index $n$. 
The band dispersion and Berry curvature can be expanded as
\begin{equation}
    \varepsilon_{\bm k} = \sum_{\bm R} \varepsilon_{\bm R} e^{i \bm k \cdot \bm R}, \qquad
    \Omega_{\bm k} = \sum_{\bm R} \Omega_{\bm R} e^{i \bm k \cdot \bm R},
\end{equation}
where
\begin{equation}
    \varepsilon_{\bm R} = V_c \int_{\bm k} \varepsilon_{\bm k} e^{-i \bm k \cdot \bm R} , \qquad
    \Omega_{\bm R} = V_c \int_{\bm k} \Omega_{\bm k} e^{-i \bm k \cdot \bm R}, \qquad \int_{\bm k} = \int_\text{BZ} \frac{d^2\bm k}{\left( 2 \pi \right)^2}.
\end{equation}

\section{Boltzmann transport equation}

The Boltzmann equation for the distribution function $f(\bm k, \bm r, t)$ in the relaxation-time approximation, is given by
\begin{equation} \label{eq:boltzmann}
    \frac{\partial f}{\partial t} + \frac{d\bm k}{dt} \cdot \nabla_{\bm k} f + \frac{d \bm r}{dt} \cdot \nabla_{\bm r} f = \frac{f^0 - f}{\tau},
\end{equation}
where $f^0(\bm k)$ is the equilibrium distribution function, i.e., $f^0(\bm k) = n_F(\varepsilon_{\bm k} - \mu)$ with $n_F(z) = 1 / (e^{z/k_B T} + 1)$ the Fermi function, with $\mu$ the chemical potential and $T$ the temperature. 

Let us consider a general uniform time-dependent electric field $\bm E(t)$. We are interested in the steady-state solutions (not necessarily static) of
\begin{equation}
    \frac{\partial f}{\partial t} - \frac{e}{\hbar} \bm E(t) \cdot \nabla_{\bm k} f = \frac{f^0 - f}{\tau}.
\end{equation}
In a translational-invariant system,
\begin{equation}
    f(\bm k, t) = \sum_{\bm R} f_{\bm R}(t) \, e^{i \bm k \cdot \bm R},
\end{equation}
where $\bm R$ are lattice vectors, and similarly for $f^0(\bm k)$. We then obtain an ordinary differential equation for each Fourier component $f_{\bm R}(t)$,
\begin{equation}
    \frac{df_{\bm R}}{dt} + \left( \frac{1}{\tau} - \frac{ie}{\hbar} \bm E(t) \cdot \bm R \right) f_{\bm R}(t) = \frac{f^0_{\bm R}}{\tau},
\end{equation}
whose general solution is given by
\begin{equation}
    f_{\bm R}(t) = \frac{f^0_{\bm R}}{\tau} \int_{t_0}^t dt' \, e^{- \frac{t - t'}{\tau}} \exp \left[ \frac{ie}{\hbar} \int_{t'}^t du \, \bm E(u) \cdot \bm R \right],
\end{equation}
with $t_0$ an integration constant.
In the static limit, i.e., for a time-independent electric field, we have 
\begin{equation}
    \lim_{\bm E(t) \rightarrow \bm E} f_{\bm R}(t) = \frac{f^0_{\bm R}}{\tau} \int_{t_0}^t dt' \, e^{-(t - t') \left( \frac{1}{\tau} - ie \bm E \cdot \bm R / \hbar \right)} = f^0_{\bm R} \, \frac{1 - e^{- \frac{t - t_0}{\tau} \left( 1 - i e \tau \bm E \cdot \bm R / \hbar \right)}}{1 - i e \tau \bm E \cdot \bm R / \hbar}.
\end{equation}
The steady-state solution is thus given by
\begin{align}
    f_{\bm R}(t) & = \frac{f^0_{\bm R}}{\tau} \int_{-\infty}^t dt' \, e^{- \frac{t - t'}{\tau}} \exp \left[ \frac{ie}{\hbar} \int_{t'}^t du \, \bm E(u) \cdot \bm R \right] \\
    & = f^0_{\bm R} \int_0^\infty ds \, e^{-s} \exp \left[ \frac{ie}{\hbar} \int_{t-s\tau}^t dt' \, \bm E(t') \cdot \bm R \right],
\end{align}
where $s = (t - t')/\tau$. The exponential factor can be interpreted as the integrated momentum shift between two scattering events at $t-s\tau$ and $t$. Going back to momentum space, we have \cite{Mikhailov2017}
\begin{equation}
    f(\bm k, t) = \sum_{\bm R} f_{\bm R}(t) e^{i \bm k \cdot \bm R} = \int_0^\infty ds \, e^{-s} f^0 \left( \bm k + \frac{e}{\hbar} \int_{t-s\tau}^t dt' \, \bm E(t') \right).
\end{equation}

We now consider the following driving field:
\begin{equation}
    \bm E(t) = \bm E_0 + \bm E_1(t), \qquad \bm E_1(t) = \bm{\mathcal E}_1 e^{i\omega t} + \bm{\mathcal E}_1^* e^{-i\omega t} = 2 \, \text{Re} \left( \bm{\mathcal E}_1 e^{i\omega t} \right),
\end{equation}
where $E_0 = |\bm E_0|$ is large compared to $E_1=|\bm E_1|$. In this case, the Fourier components of the distribution function become
\begin{align}
    f_{\bm R}(t) & = f^0_{\bm R} \int_0^\infty ds \, e^{- \left( 1 - \frac{ie \tau }{\hbar} \, \bm E_0 \cdot \bm R \right) s} \exp \left[ \frac{ie}{\hbar} \int_{t-s\tau}^t dt' \, \bm E_1(t') \cdot \bm R \right] \\
    & = f^0_{\bm R} \int_0^\infty ds \, e^{- \left( 1 - \frac{ie \tau }{\hbar} \, \bm E_0 \cdot \bm R \right) s} \exp \left[ \frac{ie}{\hbar} \left( \frac{\bm{\mathcal E}_1 \cdot \bm R}{i\omega} \, e^{i \omega t} \left( 1 - e^{-i\omega s \tau} \right) + \text{c.c.} \right) \right].
\end{align}
Up to first order in $E_1$, we can expand this as
\begin{align}
    f_{\bm R}(t) & \simeq \frac{f^0_{\bm R}}{1 - i e \tau  \bm E_0 \cdot \bm R / \hbar} + \frac{ie f^0_{\bm R}}{2\hbar} \int_0^\infty ds \, e^{- \left( 1 - \frac{ie \tau}{\hbar} \, \bm E_0 \cdot \bm R \right) s} \left[ \frac{\bm{\mathcal E}_1 \cdot \bm R}{i\omega} \, e^{i \omega t} \left( 1 - e^{-i\omega s \tau} \right) + \text{c.c.} \right] \\
    & = \frac{f^0_{\bm R}}{1 - i e \tau  \bm E_0 \cdot \bm R / \hbar} \left[ 1 + \frac{e \bm{\mathcal E}_1 \cdot \bm R / \hbar}{\omega - e \bm E_0 \cdot \bm R / \hbar - \frac{i}{\tau}} \, e^{i \omega t} - \frac{e \bm{\mathcal E}_1^* \cdot \bm R / \hbar}{\omega + e \bm E_0 \cdot \bm R / \hbar + \frac{i}{\tau}} \, e^{-i\omega t} \right].
\end{align}
Defining the frequency-space Fourier components as
\begin{equation}
    f_{\bm R}(t) = \sum_{m=-\infty}^\infty f_{\bm R,m} \, e^{im\omega t}, \qquad f_{\bm R,m}(\omega) = \frac{\omega}{2\pi} \int_0^{2\pi/\omega} dt \, f_{\bm R}(t) \, e^{-im\omega t},
\end{equation}
with $f_{\bm R,-m} = f_{-\bm R,m}^*$, we have, for example,
\begin{align}
    f_{\bm R,0} & = \frac{f^0_{\bm R}}{1 - i e \tau \bm E_0 \cdot \bm R / \hbar} \left[ 1 - \frac{2 | e \bm{\mathcal E}_1 \cdot \bm R / \hbar |^2}{\left( \omega - e \bm E_0 \cdot \bm R / \hbar - \frac{i}{\tau} \right) \left( \omega + e \bm E_0 \cdot \bm R / \hbar + \frac{i}{\tau} \right)} + \mathcal O(E_1^4) \right], \\
    f_{\bm R,1} & = \frac{f^0_{\bm R}}{1 - i e \tau  \bm E_0 \cdot \bm R / \hbar} \left[ \frac{e \bm{\mathcal E}_1 \cdot \bm R / \hbar}{\omega - e \bm E_0 \cdot \bm R / \hbar - \frac{i}{\tau}} + \mathcal O(E_1^3) \right], \\
    f_{\bm R,2} & = \frac{f^0_{\bm R}}{1 - i e \tau  \bm E_0 \cdot \bm R / \hbar} \left[ \frac{e \bm{\mathcal E}_1 \cdot \bm R / \hbar}{\omega - e \bm E_0 \cdot \bm R / \hbar - \frac{i}{\tau}} \frac{e \bm{\mathcal E}_1 \cdot \bm R / \hbar}{2\omega - e \bm E_0 \cdot \bm R / \hbar - \frac{i}{\tau}} + \mathcal O(E_1^4) \right].
\end{align}

\section{Dressed optical conductivity}

The steady-state current is given by $\bm j(t) = \bm j_\text{Bloch}(t) + \bm j_\text{geom}(t)$ with
\begin{align}
    \bm j_\text{Bloch}(t) & = - e \int_{\bm k} f_{\bm k}(t) \nabla_{\bm k} \varepsilon_{\bm k} = \frac{ie}{\hbar V_c} \sum_{\bm R} \bm R \, \varepsilon_{-\bm R} f_{\bm R}(t), \\
    \bm j_\text{geom}(t) & = -\frac{e^2}{\hbar} \left[ \bm E(t) \times \hat z \right] \int_{\bm k} f_{\bm k}(t) \Omega_{\bm k} = \frac{e^2}{\hbar V_c} \, \hat z \times \sum_{\bm R} \Omega_{-\bm R} \bm E(t) f_{\bm R}(t),
\end{align}
with
\begin{equation}
    V_c \int_{\bm k} e^{i \bm k \cdot \bm R} =  \delta_{\bm R,\bm 0}, \qquad \sum_{\bm R} e^{i \bm k \cdot \bm R} = \frac{( 2 \pi )^2}{V_c} \, \delta( \bm k ).
\end{equation}
The frequency components of the currents are thus given by
\begin{align}
    \bm j_\text{Bloch}^{(m)}(\omega) & = \frac{ie}{\hbar V_c} \sum_{\bm R} \bm R \, \varepsilon_{-\bm R} f_{\bm R,m}, \\
    \bm j_\text{geom}^{(m)}(\omega) & = \frac{e^2}{\hbar V_c} \, \hat z \times \sum_{\bm R} \Omega_{-\bm R} \left( \bm E_0 f_{\bm R,m} + \bm{\mathcal E}_1 \, f_{\bm R,m-1} + \bm{\mathcal E}_1^* \, f_{\bm R,m+1} \right),
\end{align}
with $\bm j^{(-m)} = \left( \bm j^{(m)} \right)^*$. For example, the DC component of the geometric current becomes
\begin{align}
    \bm j_\text{geom}^{(0)}(\omega) & = \frac{e^2}{\hbar V_c} \, \hat z \times \sum_{\bm R} \frac{\Omega_{-\bm R} f^0_{\bm R}}{1 - i e \tau \bm E_0 \cdot \bm R / \hbar} \Bigg\{ \left[ 1 - \frac{| e \bm{\mathcal E}_1 \cdot \bm R / \hbar |^2}{\left( \omega - e \bm E_0 \cdot \bm R / \hbar - \frac{i}{\tau} \right) \left( \omega + e \bm E_0 \cdot \bm R / \hbar + \frac{i}{\tau} \right)} \right] \bm E_0 \\
    & + \left[ \frac{\left( e \bm{\mathcal E}_1 \cdot \bm R / \hbar \right) \bm{\mathcal E}_1^*}{\omega - e \bm E_0 \cdot \bm R / \hbar - \frac{i}{\tau}} -  \frac{\left( e \bm{\mathcal E}_1^* \cdot \bm R / \hbar \right) \bm{\mathcal E}_1}{\omega + e \bm E_0 \cdot \bm R / \hbar + \frac{i}{\tau}} \right] + \mathcal O(E_1^4) \Bigg\}.
\end{align}
In lowest order of $| e \bm{\mathcal E}_1 \cdot \bm R / \hbar \omega|$, the first harmonics are given by
\begin{align}
    \bm j_\text{Bloch}^{(1)}(\omega) & = \frac{-ie}{\hbar V_c} \sum_{\bm R} \frac{\bm R\varepsilon_{-\bm R} f^0_{\bm R}}{1 - i e \tau \bm E_0 \cdot \bm R / \hbar} \frac{e \bm{\mathcal E}_1 \cdot \bm R / \hbar}{\omega + e \bm E_0 \cdot \bm R / \hbar + \frac{i}{\tau}}, \\
    \bm j_\text{geom}^{(1)}(\omega) & = \frac{e^2}{\hbar V_c} \, \hat z \times \sum_{\bm R} \frac{\Omega_{-\bm R} f^0_{\bm R}}{1 - i e \tau \bm E_0 \cdot \bm R / \hbar} \left[ \bm{\mathcal E}_1 + \frac{\left( e \bm{\mathcal E}_1 \cdot \bm R / \hbar \right) \bm E_0}{\omega - e \bm E_0 \cdot \bm R / \hbar - \frac{i}{\tau}} \right].
\end{align}
We now define the \emph{dressed} optical conductivity $\sigma_{ij}(\omega,\bm E_0)$ through $j_i^{(1)} = \sigma_{ij} \mathcal E_{1j}$. The current can thus be written as
\begin{equation}
    j_i(t) = j_i^{(0)} + 2\,\text{Re} \left[ \sigma_{ij}(\omega,\bm E_0) \mathcal E_{1j} e^{i \omega t} \right] + \mathcal O(E_1^2).
\end{equation}
Making use of $( \bm a \times \bm b)_i = \epsilon_{ijk} a^j b^k$ with $\epsilon_{i3k} = -\epsilon_{ik}$ the permutation symbol, we find\begin{align}
    \sigma_{ij}(\omega,\bm E_0) & = \frac{ie^2}{\hbar^2 V_c} \sum_{\bm R} \frac{R_i R_j \varepsilon_{-\bm R} f^0_{\bm R}}{\left( 1 - i e \tau \bm E_0 \cdot \bm R / \hbar \right) \left( \omega - e \bm E_0 \cdot \bm R / \hbar - \frac{i}{\tau} \right)} \\
    & - \frac{\epsilon_{ik} e^2}{\hbar V_c} \sum_{\bm R} \frac{\Omega_{-\bm R} f^0_{\bm R}}{1 - i e \tau \bm E_0 \cdot \bm R / \hbar} \left[ \delta_{kj} + \frac{e E_{0k} R_j / \hbar}{\omega - e \bm E_0 \cdot \bm R / \hbar - \frac{i}{\tau}} \right],
\end{align}
with $\sigma_{ij}(\omega,\bm E_0) = \sigma_{ij}(-\omega,\bm E_0)^*$. As a check, we undress the conductivity:
\begin{align}
    \sigma_{ij}(\omega,\bm 0) & = \frac{ie^2}{\hbar^2 V_c} \sum_{\bm R} \frac{R_i R_j \varepsilon_{-\bm R} f^0_{\bm R}}{\omega - \frac{i}{\tau}} - \epsilon_{ij} \frac{e^2}{\hbar} \frac{1}{V_c} \sum_{\bm R} \Omega_{-\bm R} f^0_{\bm R} \\
    & = \frac{e^2}{\hbar^2} \int_{\bm k} f^0_{\bm k} \, \frac{\partial_i \partial_j \varepsilon_{\bm k}}{i\omega + \frac{1}{\tau}} - \epsilon_{ij} \frac{e^2}{\hbar} \int_{\bm k} f^0_{\bm k} \Omega_{\bm k},
\end{align}
  with $\partial_i = \partial / \partial k_i$. This is a well-known result for the conductivity: the first term is the Drude contribution and the second term is the anomalous Hall conductivity. 

Let us now consider the dressed longitudinal conductivity $\sigma_L = \left( \sigma_{xx} + \sigma_{yy} \right) / 2$  which is a scalar and the dressed Hall conductivity, $\sigma_H = \left( \sigma_{xy} - \sigma_{yx} \right) / 2$ which is a pseudoscalar. We find
\begin{align}
    \sigma_L(\omega,\bm E_0) & = \frac{ie^2}{2\hbar^2 V_c} \sum_{\bm R} \frac{R^2 \varepsilon_{-\bm R} f^0_{\bm R}}{\left( 1 - i e \tau \bm E_0 \cdot \bm R / \hbar \right) \left( \omega - e \bm E_0 \cdot \bm R / \hbar - \frac{i}{\tau} \right)}, \\
    \sigma_H(\omega, \bm E_0) & = \frac{1}{2} \epsilon_{ij} \sigma_{ij}
    = -\frac{e^2}{\hbar} \frac{1}{V_c} \sum_{\bm R} \frac{\Omega_{-\bm R} f^0_{\bm R}}{1 - i e \tau \bm E_0 \cdot \bm R / \hbar} \left( 1 + \frac{1}{2} \frac{e \bm E_0 \cdot \bm R / \hbar}{\omega - e \bm E_0 \cdot \bm R / \hbar - \frac{i}{\tau}} \right),
\end{align}
where we used $\epsilon_{ij} \epsilon_{ik} = \delta_{jk}$. When the static field is strong, i.e, for $\omega_B \tau \gg 1$, where $\omega_B = eE_0 L/\hbar$ is the Bloch frequency with $L$ the lattice constant, $\sigma_L$ and $\sigma_H$ simplify to
\begin{align}
    \sigma_L(\omega,\bm E_0) & = -\frac{e^2}{h} \frac{\pi}{\tau \hbar V_c} \sum_{\bm R} \frac{R^2 \varepsilon_{-\bm R} f^0_{\bm R}}{\omega_{\bm R} \left( \omega - \omega_{\bm R} - \frac{i}{\tau} \right)} = \frac{e^2}{h} \frac{\pi}{V_c} \sum_{\bm R} \left( \frac{-R^2 \varepsilon_{-\bm R} f^0_{\bm R}}{\hbar \omega_{\bm R}} \right) \frac{\left( \omega - \omega_{\bm R} \right) \tau + i}{\left( \omega - \omega_{\bm R} \right)^2 \tau^2 + 1}, \\
    \sigma_H(\omega, \bm E_0) & = \frac{e^2}{h} \frac{\pi}{i\tau V_c} \sum_{\bm R} \frac{\Omega_{-\bm R} f^0_{\bm R}}{\omega - \omega_{\bm R} - \frac{i}{\tau}} = \frac{e^2}{h} \frac{\pi}{V_c} \sum_{\bm R} \left( -i\Omega_{-\bm R} f^0_{\bm R} \right) \frac{\left( \omega - \omega_{\bm R} \right) \tau + i}{\left( \omega - \omega_{\bm R} \right)^2 \tau^2 + 1} ,
\end{align}
with $\omega_{\bm R} = e \bm E_0 \cdot \bm R / \hbar$. As mentioned in the main text, we see that at resonance, $\sigma_L$ is purely reactive, while $\sigma_H$ is purely dissipative (since $\Omega_{\bm R}$ is imaginary for a system with time-reversal symmetry). Indeed, the dissipated power from the oscillating field over one period can be written as
\begin{align}
    W & = \frac{\omega}{2\pi} \int_0^{2\pi/\omega} dt \, j_i(t) E_{1i}(t) \\
    & = \frac{\omega}{2\pi} \sum_{m=-\infty}^\infty \int_0^{2\pi/\omega} dt \, j_i^{(m)}(\omega) e^{-im\omega t} \left( \mathcal E_{1i} e^{-i\omega t} + \mathcal E_{1i}^* e^{i\omega t} \right) \\
    & = j_i^{(-1)}(\omega) \mathcal E_{1i} + j_i^{(1)}(\omega) \mathcal E_{1i}^* \\
    & = \frac{\sigma_{ij}(\omega) + \sigma_{ji}^*(\omega)}{2} \, 2 \mathcal E_{1i}^* \mathcal E_{1j} \\
    & = \frac{\sigma_{ij}(\omega) + \sigma_{ji}^*(\omega)}{2} \, \mathcal E_{1i}^* \mathcal E_{1j} + \frac{\sigma_{ij}(-\omega) + \sigma_{ji}^*(-\omega)}{2} \, \mathcal E_{1i} \mathcal E_{1j}^* 
\end{align}
Hence the absorpative part of the conductivity tensor is given by the Hermitian part:
\begin{equation}
    \sigma^\text{abs}_{ij} = \text{Re} \, \frac{\sigma_{ij} + \sigma_{ji}}{2} 
    + i \, \text{Im} \, \frac{\sigma_{ij} - \sigma_{ji}}{2} = \text{Re} \, \frac{\sigma_{ij} + \sigma_{ji}}{2} + i \epsilon_{ij} \, \text{Im} \, \sigma_H.
\end{equation}
Similarly, the reactive part of the conductivity tensor is given by the anti-Hermitian part. For linearly polarized light, we can take $\mathcal E_1$ real and $\text{Im} \, \sigma_H$ does not contribute to dissipation. However, this term does give rise to dissipation for circularly polarized light. In this case, the imaginary part of the optical Hall conductivity gives a transverse response that lags by a quarter cycle. Hence for circular polarization $\mathcal E_{1x} = \pm i \mathcal E_{1y}$, the current response due to $\text{Im} \, \sigma_H$ actually lies parallel to the field and contributes to dissipation.

\end{document}